# Charging effects and anomalous resistive features of superconducting boron doped diamond films.

Christopher Coleman[1] and Somnath Bhattacharyya [1,2,(a)]

[1]*Nano-Scale Transport Physics Laboratory, School of Physics and DST/NRF Centre of Excellence in Strong materials, University of the Witwatersrand, Johannesburg 2050, South Africa*

[2]*National Institute of Science and Technology "MISiS", 119049, Leninsky pr. 4, Moscow, Russia*

**Abstract** – Anomalous resistive peaks below the superconducting transition temperature in heavily boron doped nanocrystalline diamond films could have potential application in switching devices, however it's exact origin is still under study. We establish a temperature dependence of this resistive phase similar to what has been reported for in Josephson junction arrays and other granular superconductors where the charge duel of the Berezinskii-Kosterlitz-Thouless (BKT) transition has been observed. Non-linear magnetoresistance with a temperature dependent peak feature below the critical field are also presented. Pronounced temperature dependent hysteresis in the current voltage sweeps at temperatures below the determined BKT critical point are related to pinning of charge defects. It is shown that these collective features allude to a Charge-BKT transition between charge and anti-charge analogues.

**Introduction**

Superconducting Josephson Junction arrays have long been known to exhibit several interesting phenomena important for technological application. This includes a vast range of technologies including superconducting magnetometrers, space detectors and Qubits. Through changing the charging to Josephson energy ratio ($E_C/E_J$), it is possible to switch between a low and high resistive regime[1]. What makes these systems even more interesting is that the difference between these two regimes can be explained through the concept of duality[2]; on the superconducting side ($E_C < E_J$) Josephson tunnelling dominates and cooper pairs are free to conduct, on the insulating side ($E_C > E_J$) Cooper pairs are localized leading to a resistive phase generally referred to as a bosonic insulator[3] (which is essentially a charge - anti charge plasma).

In systems with high tunnelling resistance, excess charge can build up in respective arrays or grains due to single particle tunnelling, this in turn leads to charging effects where excess charge in an island/electrode can polarize neighbouring electrodes[4]. This has been experimentally related to the formation of charge solitons[5] which can dominate the transport properties. In 2D arrays when the distance between two charge solitons is smaller than the screening length yet larger than the thickness ($d < r \ll \Lambda$, figure 1 a) the interaction between the charge anti-charge pair is essentially logarithmic. This leads to the observation of the charge analogue of the BKT transition (c-BKT)[6]. As the bound pairs are electrically neutral the system will be in a resistive state even at temperatures below the critical point. These features have also been observed in granular superconducting systems where superconducting islands are separated by fine grain boundaries acting as Josephson junctions[7].

One system of particular interest is the nanocrystalline boron doped diamond, not only has this system already exhibit a pronounced re-entrant resistive phase[8,9,10] but also effects of cooper pair confinement[11], low dimensionality[12,13] and signatures of a BKT like transition[14]. The boron doped diamond system is considered a structural granular superconductor composed of nano-scale crystalline diamond grains. High resolution transmission electron microscopy indicates columnar growth with crystal twinning between grains, thus the microstructure of the system is more ordered than conventional granular superconducting films (usually grown through sputtering or other deposition techniques) and due to the insulating nature of intrinsic band structure of diamond still present in the system presents an ideal system to study charging effects mentioned above. One of the most intriguing aspects of this system is the formation of an anomalous resistive phase, the bosonic insulator.

In earlier works we have demonstrated that this resistive phase does show temperature dependent scaling characteristic of a Charge-Kondo effect[10]. Thus in analogy to a Josephson array where the increase in charging energy leads to a resistive state, the boron doped diamond films, boron dopants can act as degenerate Kondo impurities by opening pseudo-spin (charge) scattering channels and cause the observed resistive phase. Furthermore, the existence of such charge impurities in a low dimensional superconducting system has huge implications on the transport of the system and are ideal nucleation sites or pinning centres for the formation of charge-solitons. Furthermore, recent theoretical investigations[15] have shown that many body correlations in Josephson junction arrays can lead charge-polarization and hence a Josephson-Kondo screening effect, thus reconciling the occurrence of a Charge-Kondo effect with soliton formation in a superconducting system.

In this work we further demonstrate the link between charging effects and the tunability of the bosonic insulating phase of boron doped diamond. It is observed that the measurement bias has an upper bound for the observation of the resistive peak. This dependence on the current strength is related to the self-capacitance of the nano-scaled grains and thus the screening length (as in the case of a Kondo effect) of trapped charges. The applied bias thus offers an experimental means of tuning between the superconducting and resistive regimes. Furthermore, we investigate the magnetoresistance of this system at different measurement bias, we observed a peak in the magnetoresistance at field strength below the critical field. The peak amplitude and

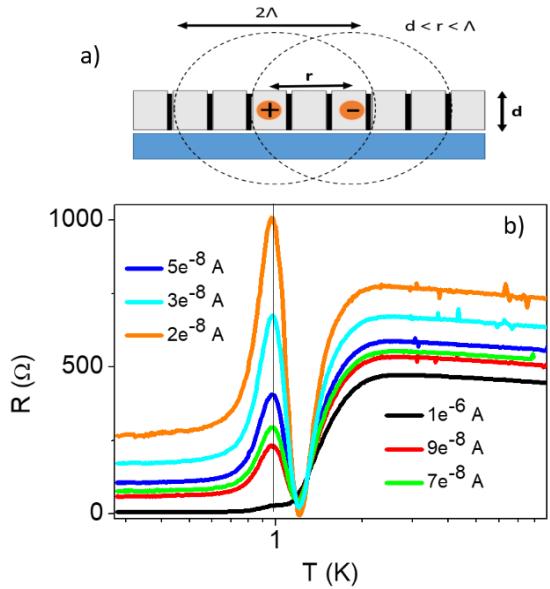

*Figure 1 a) Schematic showing the dimensional considerations for observing a quasi-2D interaction between localized solitons in a Josephson Junction array. b) The boron doped diamond films show a tuneable transition between complete superconductivity for measurement bias above 1 µA and a pronounced insulating peak for bias below this value.*

position show a dependence on the measurement bias as well as temperature. These features are compared to similar phenomena observed in field and disorder induced superconductor to insulator transitions previously reported[16]. Finally, we present a device concept based on the tunability of the resistive phase in boron doped diamond, relating the controllable binding of charge solitons to entangled states forming a collective multilevel system that may be of use for quantum information processing recently established in other soliton systems.

**Results**

Superconducting boron doped diamond samples were grown using a microwave plasm enhanced chemical vapour deposition technique. Trimethyl-boron, hydrogen and Argon are used as precursor gasses. The methane to hydrogen ration is particularly important for controlling the structural properties such as average grain size[17]. The boron concentration was determined to be up to 3 %$_{at}$ as required for observing the superconducting phase. SEM analysis and HR-TEM are used to determine the average grain size as well as microstructure of the system. Electrical measurements are conducted at temperatures between 300 mK and 10 K to investigate the transition point and re-entrant resistive phase in the fluctuation regime. As shown in figure 1 b. the boron doped diamond system can be tuned from a complete superconducting state to one showing a resistive peak and saturating finite resistance through lowering the measurement bias. At measurement bias below ~$1 \times 10^{-7}$ A the resistive peak height increases drastically as the measurement bias is decreased. Previous results[14] have shown the IV characteristics scale according to a power law

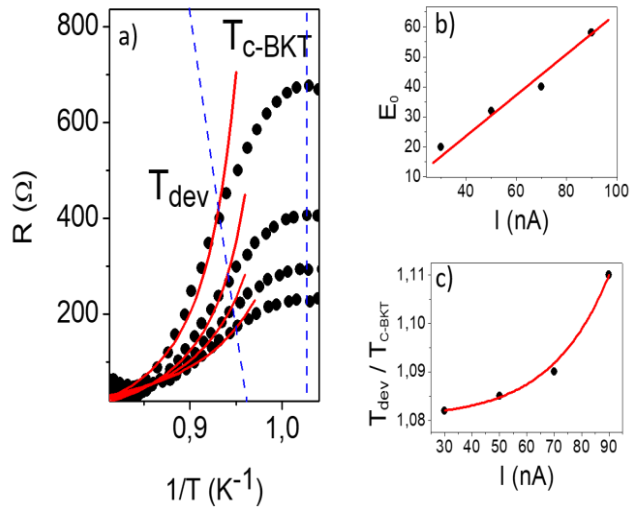

*Figure 2 a) The resistive peaks have been fitted to equation 1, and yielding $T_{dev}$ and $T_{c-bkt}$ values, as indicated by the dashed lines b) The $E_0$ fitting parameter is observed to follow a linear dependence on the measurement current, as expected. c) $T_{dev}/T_{c-bkt}$ ratio is observed to increase exponentially with increasing bias, this is an indication of suppression of the screening length.*

indicative of a BKT transition, following convention to further demonstrate the transition here the resistive upturns have been fitted to the square-root-cusp relationship[4,5,6]:

$$R(T) = E_0(I)\exp\left(\frac{2b}{\sqrt{T/T_{c-bkt}-1}}\right), \qquad (1)$$

Where $E_0(I)$ is a fitting parameter related to the normal state resistance and is thus dependent on the applied current and dimensions of the junctions (grain boundaries), b is a constant or order ~ 1 and $T_{c\text{-}bkt}$ is the determined c-BKT transition temperature (which occurs at the peak height as determined before[14]). As shown in figure 2 a, the upturn in the temperature dependent resistance can be fitted to this formula up until the characteristic deviation temperature denoted $T_{dev}$. This feature has been observed n fabricated arrays driven into the resistive regime[5] and is an indirect indication of the screening length. Following the analysis of ref 5, as the $T_{dev}/T_{c\text{-}bkt}$ ratio increases the screening length is expected to decrease. As this decrease in $T_{dev}$ is tuned by the applied bias it is natural to correlate the measurement bias to the screening length ($\Lambda$). This is quite an interesting finding as the screening length is defined by $\Lambda = \sqrt{C/C_0}$, where C is the capacitance of the junctions and $C_0$ the self-capacitance of the individual grains, It is known that the self-capacitance of nano-structures are high dependent on the density of charge carriers[17]. Thus the measurement bias can indeed alter the screening length of localized charges through changes in the self-capacitance arising from tunnelling of charge carriers between grains. This effect becomes particularly important at temperatures deep within the fluctuation regime where finite number of normal state charge carriers dominate the transport properties in granular superconducting systems. Qualitatively this means that at lower bias (where the resistant peak is most prominent) a larger screening length is obtained, this fulfils the strict dimensionality requirements (d < r << Λ) for the observation of the c-BKT transition and thus a larger peak height is observed, upon increasing the bias strength, the screening length decreases until it is comparable to or smaller than the distance between soliton charges, this signals the breakdown of the c-BKT transition and thus a complete suppression of the resistance peak (as observed in figure 1 b).

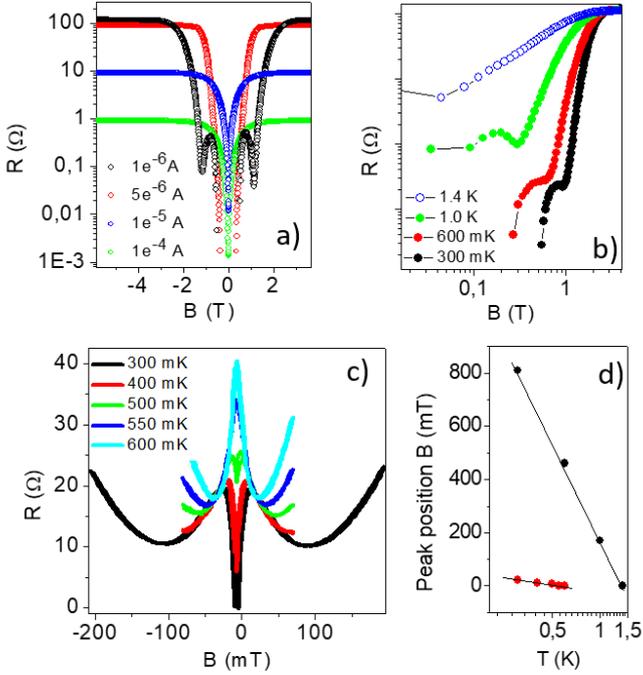

*Figure 3. a) Magnetoresistance show non-monotonic behaviour with a peak structure that occurs for the same bias strength observed in the temperature dependent resistance. b) This peak structure shows a strong temperature dependence and shifts to lower fields upon increasing temperature. c) The peak in the magnetoresistance (measured at 1 µA) is also heavily dependent on the average grain size and occurs at a much-reduced field strength for larger grains). d) The peak position occurs at almost an order of magnitude lower field strength for samples with grain size ~100 nm (red) compared to samples with average of average grain size ~30 nm (black).*

To further probe the effect of measurement bias on the resistive phase the magnetoresistance is measured under both low and high bias conditions. As shown in figure 3 a, the magnetoresistance also shows a bias dependence for the observation of peak structures in the isotherms. As with the temperature dependent resistance, a peak structure appears in the magnetoresistance only for bias below a threshold value of approximately ~1x10$^{-7}$ A, this is indicated by the black curve in figure 3 a. The magnetoresistance peaks also showed dependence on the average grain size of the samples, and as shown in figures 3 c,d, for larger grain size samples the peaks occur at a much lower magnetic field with a smaller maximum height. This can again be directly related back to the dimensional restrictions for observing a logarithmic interaction between charge anti-charge pairs. Similar non-linear magnetoresistance peak structures have recently been reported as indications of the c-BKT transition in NbTiN films[16]. There too the resistive peak structure was observed within the same field strength range and was related to the divergence of the dielectric constant (ε) as a function of the carrier concentration (applied bias in our system). The magnetoresistance peak structures observed in our samples shift to lower field strength upon increasing temperature (figure 3 b,c,d). This indicates a temperature related instability in the binding of the charge-anti charge pairs most likely due to charge carrier tunnelling between grains according to the Arrhenius activation behaviour as the temperature increases. This observation also signifies the effect of applied field on the screening length and consequently depression of the $T_{c\text{-}BKT}$ (resistive peak maximum) with increasing field strength[8,9,10]. As the applied field reduces the screening length, increasing field strength reduces the soliton interaction and thus drives the observation of a c-BKT transition to lower temperatures, manifesting as a temperature dependent peak in the magnetoresistance.

Further evidence for soliton dynamics present in the thin films is the hysteretic behaviour of the current voltage characteristics. As shown in figure 4 a, a pronounced hysteresis can be observed when sweeping between positive and negative bias. This is characteristic for systems with appreciable pinning forces[19]. Similar features have before been examined theoretically[19] for regular Josephson Junction arrays. There it was found that the ratio of the depinning current to critical current followed a distinct trend as a function of the soliton size (effective screening length). This analysis is particularly valuable in our investigation as both field and temperature are shown here to influence this length scale. From our data we extract the depinning current ($I_{dep}$) and the critical current ($I_c$) (here taken to be approximately the switching current) from the hysteretic IV curves (inset figure 4). The depinning current is found to be strongly temperature dependent and is only discernible at temperatures below the determined $T_{C-BKT}$. As shown before[18], the depinning force can at least qualitatively be related to this increase in soliton screening length. This is established in figure 4 a, where the ratio of the depinning current to critical current is plotted as a function of decreasing temperature. For temperatures below 1.3 K (approximately the temperature where the Bosonic Insulating phase appears) the depinning force neatly fits the predicted trend[19]:

$$I_{dep}/I_C = \frac{f}{\pi}\sqrt{\frac{2\eta}{a}[1-\frac{\eta}{a}]} + \alpha e^{-\beta/a}, \quad (2)$$

Valid for $\eta \leq a \leq 1$, where $\eta$ is the bias current normalized to critical current, f ~ 0.8 describes the strength of pinning force and $a$ is the so-called discreetness of the soliton and has an inverse temperature dependence. The exponential term is the Peierls-Nabarro pinning potential (where α and β are numerical parameters determined through fitting). Qualitatively what this analysis signifies is a decrease in the pinning current as the temperature is lowered, this corresponds to increasing soliton size as discussed above.

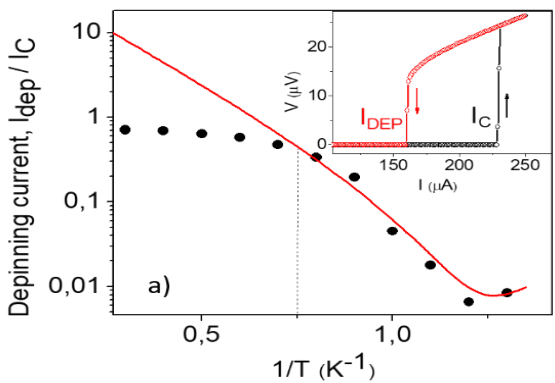

*Figure 4. a) From the hysteresis in the IV characteristics (inset) the depinning and critical currents can be determined, this ratio is found to follow the theoretically predicted trend, given by equation 2.*

Thus, dynamics such as depinning of the charge solitons can be manipulated in a controllable way and monitored through this temperature dependent hysteresis. The possibilities of charge solitons dominating the resistive phase in boron doped diamond is particularly interesting for the development of quantum technologies[20,21,22]. This is because solitons in various physical systems can display phenomena such as quasi-superradiance[23] and act as multi-level dark states[24,25] and thus can be useful for developing novel qubits or quantum memory nodes.

**Conclusion**

The collection of features presented above indicate that the previously reported[14] possibility of a BKT transition in boron doped diamond can in fact occur when considering the charge dual. This is due to the effective dimensional considerations related to the thickness, distance between charges as well as the screening length. The bias dependence on the peak height is related to the screening length of such charge solitons and explains why the peak is only observed under very specific measurement conditions (low measurement bias). Non-monotonic magnetoresistance in the form of a peak structure are also observed in the same bias regime, these features are also presented as evidence for the c-BKT effect. Furthermore, depinning effects of the solitons are observed in the IV characteristics through a pronounced temperature dependent hysteresis. Boron doped diamond thus offers an interesting system where the enhanced carrier correlations lead to effects such as a tunable resistive phase due to a charge-Kondo effect, formation of solitons in this resistive state and finally a c-BKT transition.

**Acknowledgments**

SB is very thankful to Miloš Nesládek for supplying the samples and D. Mtsuko for valuable discussion. CSIR-NLC, the URC Wits and National Research Foundation (SA) for the Nanotech Flagship Project and the BRICS multilateral program for funding.